# Extinction times of epidemic outbreaks in networks

Petter Holme[1,2,3]

[1]Department of Energy Science, Sungkyunkwan University, 440-746 Suwon, Korea
[2]IceLab, Department of Physics, Umeå University, 90187 Umeå, Sweden
[3]Department of Sociology, Stockholm University, 10961 Stockholm, Sweden

E-mail address: holme@skku.edu

Abstract

In the Susceptible–Infectious–Recovered (SIR) model of disease spreading, the time to extinction of the epidemics happens at an intermediate value of the per-contact transmission probability. Too contagious infections burn out fast in the population. Infections that are not contagious enough die out before they spread to a large fraction of the population. We characterize how the maximal extinction time in SIR simulations on networks depend on the network structure. For example we find that the average distances in isolated components, weighted by the component size, is a good predictor of the maximal time to extinction. The transmission probability giving the longest outbreaks is larger than, but otherwise seemingly independent of, the epidemic threshold.

Introduction

Over the last decades the mathematical and computational study of infectious disease epidemiology has increasingly come to focus on the structure of human contact patterns. Such structure affects disease spreading in different ways and can explain phenomena that earlier approaches (based on the assumption that any pair of individuals have the same chance of meeting at any given time) cannot. For example, the relatively slow early phase of some outbreaks have been related to clustering (many short cycles) in the contact network [1]; the existence of "super spreaders" (people infecting many more others than the average infectious person do) has been attributed to the broad distribution degree (number of neighbors) in the contact network [2]. One, we think, understudied topic of network epidemiology is the time for an epidemic outbreak to die—the *extinction time*. Not all diseases have such a life cycle of birth and extinction, of course, but some do and the canonical model for these is the Susceptible–Infectious–Recovered (SIR) model [3]. In this paper, we will study extinction times of the SIR model on various types of networks.

Research on the time-scales of epidemic processes has typically focused on the early phase of the outbreak [1,4], probably because it is mathematically simpler. Calculating the time to extinction simultaneously needs to account for the chance of the disease to die out early, and the time it would take to burn out in the population. These competing effects are hard to handle analytically, even with an approximate approach. In this paper, we study extinction times computationally. For most (perhaps all) networks, the extinction time as a function of the per-contact-transmission probability λ will have a maximum for an intermediate λ-value. This is a consequence of the mentioned competing effects of early extinction (at small λ) [5] and fast spreading speeds (at large λ). We will investigate many aspects of this maximum and how it relates to the structure of the underlying networks.

Results

In this section, we present the results for our simulations. We use three basic types of networks: Networks with a power-law degree distribution generated by the configuration model (see the Methods section). These have a probability of a vertex with degree $k$ proportional to $k^{-\gamma}$. We try different γ-values—2, 2.5 and 3—both because empirical network data often show degree exponents in that range, and the configuration model changes dramatically between these values. When γ = 2 (in the large $N$-limit) the network is almost surely connected into one big component; when γ = 2.5, the largest component is about $0.62N$, and when γ = 3, the largest component is approximately $0.19N$ [6]. The configuration model has a vanishingly small fraction of triangles (i.e. the transitivity [6] goes to zero as $N \to \infty$). Since many triangles, as mentioned above, can affect disease spreading, we also modify the configuration model network by adding triangles. We repeatedly connect two distinct random neighbors of a random vertex. In this process, the edge is added to vertices with a probability proportional to their degrees—a process ("preferential attachment") that can create networks with power-law degree distributions. On the other hand, in this setup (where no edges are deleted or vertices added) this edge adding will eventually destroy the power-law degree distribution [7]. This does not matter to us; there is nothing in our analysis that depends on the specific form of the degree distribution. In fact, we also use networks of the Erdős-Rényi model with the much narrower Poisson degree distribution (see the Methods section).

In our first analysis, we illustrate how the average extinction time $\langle \tau \rangle$ depends on the per-contact transmission probability λ. We run the SIR model on a realization of the configuration model with 200 vertices and a power-law degree distribution with exponent –2.5 (for details, see the Methods section). In Fig. 1A we see that $\langle \tau \rangle$ peaks for intermediate λ-values. For less contagious diseases (small λ), ⟨τ⟩ is low because the disease does not spread very far before it dies. For highly contagious diseases (large λ), the disease burns out fast in the population. Around λ = 0.42 the average extinction time is the longest. One may believe that the maximizing λ, $\hat{\lambda}$, goes to zero as $N$ increases, just like the epidemic threshold [8]. However, as we will show below, that does not seem to be the case. The ⟨τ⟩-peak does not coincide with any feature of the average fraction of at-some-point infected vertices Ω as a function of λ (Fig. 1B). (Ω(λ) is an indicator function for the epidemic threshold.)

In the case λ = 1, we can easily derive the time to extinction. In this case, since the disease travels maximally one step at a time, a vertex located a distance $e$ from the seed $i$ of the infection will be infected after $e$ time steps. The outbreak will thus be extinct at time $e(G_n,i) + δ$, where $G_n$ is the component where the infection seed is located. To get the estimated extinction time, we have to average over all vertices as seeds and also weigh the average $e(G_n,i)$ values by the size of $G_n$ (since the chance of the seed belonging to a certain component is proportional to its size). The resulting metric, *size weighted average eccentricity* (SWAE), is discussed further in the Methods section. For λ close to zero, the outbreaks are small and brief. We know that local network structures such as degree distribution [4] and clustering [1] are important factors for the spreading speed. The maximal time to extinction happens for intermediate λ-values and is ultimately dependent on more aspects of the network structure than either of the limit cases. We will explore the idea that the maximal time to extinction can be predicted by a combination of a distance and component-size metrics like SWAE. We test eight such measures (described in detail in the Methods section) to see which one that best predicts the maximal time to extinction.

As an example of the correlation between extinction times and measures combining distances and component sizes, we show (in Fig. 2) scatter plots of the maximal extinction time as a function of SWAE for all our model topologies. We see that SWAE does a fairly good job—better than the size of the network—to predict the maximal outbreak time. The effect of the clustering is small (comparing Fig. 2A and B)—it just lifts the point cloud a bit. In the Erdős-Rényi model case, the spread of the points is larger. Furthermore, the points from different network sizes overlap less. In general, the results for the Erdős-Rényi model fluctuate much less (as anticipated, see Ref. [9]), than the scale-free networks. For the Erdős-Rényi model, the *largest-component average distance* (LCAD) is a better predictor than SWAE (we discuss this further below).

To summarize this type of scatterplots, we perform a linear regression analysis and calculate the coefficient of determination $R^2$ as a goodness-of-fit measure. The better the fit is, the higher is the explanatory power in the cluster-size vs. distance metric. In Fig. 3, we show the results for our three types of networks as functions of the numbers of vertices. For almost all the networks with a power-law degree distribution SWAE, *size-weighted average distance* (SWAD) and *size-weighted diameter* (SWD) shows the strongest correlation. The exception is γ = 3 with added triangles where LCAD has higher $R^2$-values than SWD. Among the three size-weighted measures (where the contribution from a component is weighted by its size), SWAD is doing best as a predictor, followed by SWAE and SWD. As mentioned, at λ = 1, you just have to add δ to SWAE to get the exact average maximal extinction time. The maximizing λ-value is on the other hand far from unity. This means that other pathways than the shortest comes into play and their distances seems to be captured better by SWAD. This feels natural since SWAD incorporates more information ($N - 1$ shortest paths per vertex instead of just one for SWAE), but it is hard to give some more deductive argument, let alone an analytical one. Some of the curves in Fig. 3C–F are decaying with $N$. For γ = 2.5 and 3 networks, both the *harmonic mean distance* (HMD) and the *largest component size* (LCS) are decaying. The HMC puts a larger weight on smaller distances and depends more strongly on the number of small components. At the same time, evidently, the average maximal time to extinction depends more on the largest component (as measured by LCAD), which explain the poor performance of HMC. The largest-component based measures have the steepest rise for all the plots. For the γ = 2 figures, they will eventually reach the size-weighted measures (since these networks have only one component in the large-$N$ limit and fewer components, on average, for finite sizes); for γ = 2.5 and 3 it is not yet clear. The goodness of fit curves for the Erdős-Rényi model, Fig. 3G, looks rather different. The degree distribution does apparently influence the outcome considerably. The most predictive measures are the two average distances (LCAD and SWAD). We chose the parameter values so that it usually will exist at least one component of considerable size, but rarely all vertices would be connected into one component. Apparently, the largest component is still more influential than the others. One reason for this could be that it has a higher average degree than the rest of the network and these extra possibilities of infection that comes with the extra edges make outbreaks able to spread in the largest, but not the other components. Just as for the networks with more heterogeneous degree distributions (regardless if we consider an SW or AD measure), the diameter based measures give lower predictability than the average eccentricity ones, while the average distance measures show the highest $R^2$-values. Assuming a connected network (or a network with a converged component-size distribution), the average distances in the components of the Erdős-Rényi model scales like log $N$, and in scale-free networks like log log $N$ for $\gamma < 2$, like log $N$ / log log $N$ for $\gamma = 2$, and like log $N$ for $\gamma > 2$ [10]. We can assume $\langle \tau \rangle$ has the same scaling behavior.

We finish our analysis by a shifting our angle a bit and consider the epidemic threshold $\lambda_c$. If $\lambda < \lambda_c$, then $\Omega = 0$ in the $N \to \infty$ limit. If $\lambda > \lambda_c$, then $\Omega > 0$ (i.e., the disease can spread in the population). Networks with a power-law degree distribution and $\gamma \leq 3$ are known to give the SIR model an epidemic threshold $\lambda_c = 0$. In the canonical models of statistical physics, many quantities simultaneously show peaks or singularities at a phase transition point. Therefore, one could anticipate that the $\lambda$-value giving the maximal time to extinction, $\hat{\lambda}$, would tend to zero as the system size increases. What we observe for the scale-free networks (Fig. 4A) are some weak trends in both directions. One would need to run simulations for larger system sizes to make a conclusive statement. The trends for the clustered configuration-model networks are the same (Fig. 4B), but the values are 0.1–0.2 lower. For the Erdős-Rényi model networks, the trend is clear and decreasing. The epidemic threshold in this case is ill-defined since the Erdős-Rényi model, at the critical point (the parameter values we use), has no giant component (the property that the largest component scales like $N$).

## Discussion

We have investigated a characteristic time of the SIR disease-spreading model; namely the time to extinction. Given a static contact network, this quantity has a maximum for intermediate values of the per-contact transmission probability. The reason is that for low transmission probabilities, the chance for the disease to die out early is higher; for high transmission probabilities, the speed of the infection front is faster. The maximal time to extinction is an interesting quantity, complementing the often-studied final outbreak size. A long time-to-extinction gives the society longer time to mobilize counter measures. We find that the component sizes and the distances within the components can rather accurately predict the maximal time to extinction. Of the various measures combining component sizes and distances are those using average distances performing consistently better. This is a bit surprising since the extinction of the outbreak is an extreme event, and that would presumably be more correlated to extreme distances such as eccentricity. Indeed, in the limit of high transmission probabilities, the extinction time would be exactly proportional to the average eccentricity weighted by the size of the components. Our different measures use different ways of dealing with the component sizes—they either weigh results by the size of the components or use only the largest component. For networks with heterogeneous degree distributions, the size-weighted measures perform better; for the Erdős-Rényi model networks, the largest-component based measures perform better. One explanation could be that for the Erdős-Rényi model networks, the average degrees in the non-largest components are so low that they are under the epidemic threshold, and thus would not contribute much to the extinction times. Another conclusion we make is that the maximum of the extinction time is unrelated to the epidemic threshold. This means that the effect of fast burnouts (lowering the extinction times) only becomes an important factor well inside the parameter region of epidemic spreading.

When it comes to real epidemics, this research probably applies best to pathogens spreading via sexual contacts [11] because the contacts are rather well defined and their network structure (with heterogeneous degree distributions) have a strong influence on sexually transmittable epidemics.

We hope this work can inspire more research in the long-time behavior of epidemic models. It would be natural to extend this work to temporal networks [12] or models with coevolving awareness of the outbreak [13].

## Methods

### Disease simulation

We use an individual-based SIR model. Each individual belongs to one of three states Susceptible (S), Infectious (I) or Recovered (R). The $N$ individuals are connected into a network $G = (V,E)$ where $V$ is the set of individuals (or vertices) and $E$ is the set of connected pairs (edges) of vertices. Time proceeds in discrete time steps. If $i$ is a Susceptible individual, then at every time step, an edge to an Infectious neighbor represents a potential infection event. At such an event, $i$ becomes Infectious the next time step with a probability $\lambda$. The simulation starts with all vertices being S, except a random individual turning I at time $t = 0$. An Infectious individual turns Recovered $\delta$ time steps after it became Infectious. In this work we use $\delta = 4$ (we briefly test $\delta = 3, 5, 10$ and get the same conclusions). Recovered vertices stay Recovered for the rest of the simulation. The simulation ends when none of the individuals is Infectious. The output we measure from a run is the time to extinction $\tau$ and the fraction of individuals at some time infected.

### Measuring the maximal time to extinction

Given a graph $G$, the expectation value of $\tau$, $\langle\tau\rangle$, is a unimodal function of $\lambda$. To find its maximum, we use the "bounded" method of the "optimize.minimize_scalar" function of the Python 2.7.8 package SciPy (http://www.scipy.org/). This function uses a hybrid between golden ratio search and Brent's method to locate the maximum in fewest possible measure-ments of $\langle\tau\rangle(G,\lambda)$ (typically around eight measurements are needed for the sizes in this paper). To calculate $\langle\tau\rangle(G,\lambda)$, we iterate at least 100 (in practice $10^5$–$10^6$ times) until the standard error of $\langle\tau\rangle$ is less than 0.1% of $\langle\tau\rangle$.

**Network generation**

We base our network generation on the configuration model [14]. In other words, we draw $N$ numbers, $k_1, \ldots, k_N$, from a probability distribution $p(k)$. In our case, we use the discrete distribution with probability mass function

$$p(k) = \begin{cases} ck^{-\alpha} & \text{for } 1 \leq k \leq N \\ 0 & \text{otherwise} \end{cases}.$$

(1)

Here $c$ is a normalization constant. This is an emergent power-law distribution. The purpose with the upper bound ($k \leq N$) is to avoid extremely high degrees (which would consume much memory in the simulations). The results in the large-$N$ limit and the qualitative observations are the same as for the more common (especially in theoretical calculations) formulation without the upper bound. Note that the generated graphs are multigraphs where both multiple edges and self-edges are allowed.

Let $M$ be the number of added edges in the configuration model as described above. We test the effect of triangles by adding such to the configuration model networks. We randomly pick two distinct neighbors of a random vertex and add an edge (whether or not there is a link between these two vertices before). This is repeated until $\eta M$ triangles are added. In this paper we use $\eta = 0.6$.

Furthermore, we use the Erdős-Rényi model [15]. In this model one start from $N$ disconnected vertices and add $M$ edges between random pairs of distinct vertices that are not already connected. We use $M = N$ to get networks that are somewhat fragmented, but still having large components. Erdős-Rényi model, unlike the configuration model, creates simple graphs (without self-edges or multiple edges).

**Distance and component-size metrics**

(For a good general introduction to distance measures in graphs, see Ref. [16].) Let $G_1$, $G_2, \ldots, G_C$, be a decomposition of $G$ into $C$ *components*—disconnected subgraphs that are maximal (in the sense there one cannot connect a vertex to a component by adding an edge to it) and connected (so that there is a path between any pair of vertices of the same component). Let $d(i,j,G_n)$ be the distance (number of edges in a shortest path) between $i$ and $j$ in component $G_n$. Then we can define the following eight distance and component-size measures:

*Size weighted diameter*

$$\text{SWD}(G) = \frac{1}{N(G)} \sum_{i=1}^{C} N(G_i) \max_{i,j} d(i,j,G_i)$$

(2)

where $N(G)$ is the number of vertices in $G$. This is the expected distance if you pick a random vertex and measure the average diameter (largest distance between any pair of vertices) of its component.

*Size weighted average eccentricity*

$$\text{SWAE}(G) = \frac{1}{N(G)} \sum_{i=1}^{C} \sum_{j \in V_i} \max_k d(j,k,G_i)$$

(3)

The eccentricity of a vertex is the distance to the most distant vertex in the same component. SWAE is the expected eccentricity of a random vertex.

*Size weighted average distance*

$$\text{SWAD}(G) = \frac{2}{N(G)} \sum_{i=1}^{C} \frac{1}{N(G_i) - 1} \sum_{j < k \in V_i} d(j,k,G_i)$$

(4)

This is the expected value of the average distance between two vertices in the component of a randomly chosen vertex.

*Harmonic mean distance*

$$\text{HMD}(G) = \left[ \frac{2}{N(G)[N(G) - 1]} \sum_{j < k \in V_i} \frac{1}{d(j,k,G)} \right]^{-1}$$

(5)

This is another way to weigh together component sizes and distances within the components. It is appealing in its simplicity (it does not need an explicit enumeration of the component) [17]. On the other hand, it lacks a motivation from processes on the graph. Its reciprocal value $1/\text{HMD}(G)$ was introduced under the name "efficiency" in Ref. [18].

*Largest component diameter*

$$\text{LCD}(G) = \max_{i, \in G'} d(i,j,G')$$

(6)

where $G'$ is the largest component of $G$. The largest component determines many interesting properties such as the maximal outbreak size. Real-world networks also tend to be connected into one component (sometimes called "giant component" by analogy to graph models).

*Largest component average eccentricity*

$$\text{LCAE}(G) = \frac{1}{N(G')} \sum_{i \in G'} \max_{j \in G'} d(i,j,G') \tag{7}$$

This is the average distance from a vertex of the largest component and its most distant vertex in the same component.

*Largest component average distance*

$$\text{LCAD}(G) = \frac{2}{N(G')[N(G')-1]} \sum_{i<j \in G'} d(i,j,G') \tag{8}$$

It is the same as LCAE but averaging the distance from *i* to all other vertices rather than only the one furthest away.

*Largest component size*

$$\text{LCS}(G) = N(G') \tag{9}$$

This is simply the size of the largest component (mostly for comparison—it could clearly not predict epidemic time scales in general).

Figures

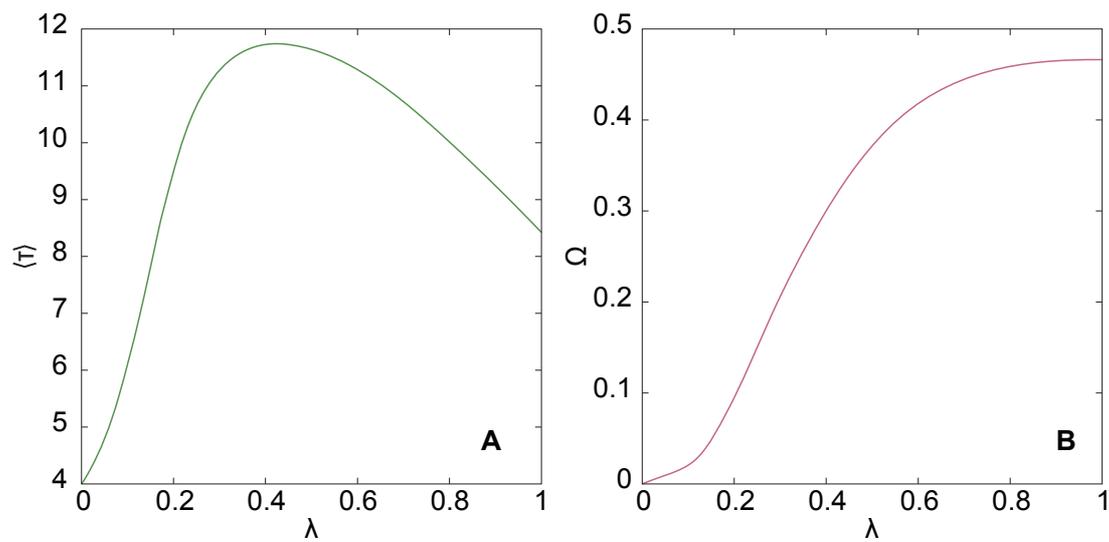

**Fig. 1.** The average time to extinction as a function of the per-contact transmission probability (A). Panel B shows the average fraction of vertices at some point affected by the outbreak. For an intermediate value $\hat{\lambda}$ the outbreak lives longer than for other $\lambda$ (A). These figures are made for one realization of the configuration model with power-law degree distribution with parameter values $N = 200$ and $\gamma = 2.5$. We sampled 501 equidistant values of $\lambda$, averaged over $10^6$ runs of the disease simulation. The curves were smoothed to remove small fluctuations (the general curvature on a scale larger than $\lambda \approx 0.001$ is unaltered).

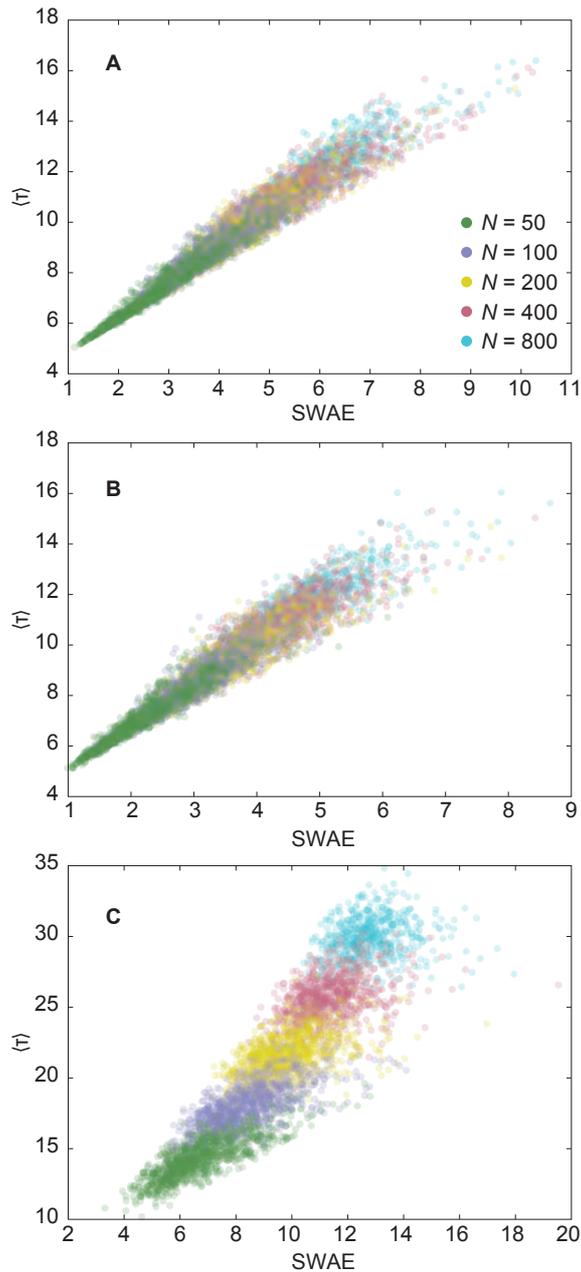

**Fig. 2.** Scatter plots of the maximal time to extinction as function of SWAE. A shows the results for the configuration model; B displays the result for the configuration model with added clustering; C shows the outcome for the Erdős-Rényi model. One dot gives the value for one realization of the network models (they have transparency to visualize the overlap of points). There are 1000 points per size. Error bars in ⟨τ⟩ would have been smaller than the symbol size.

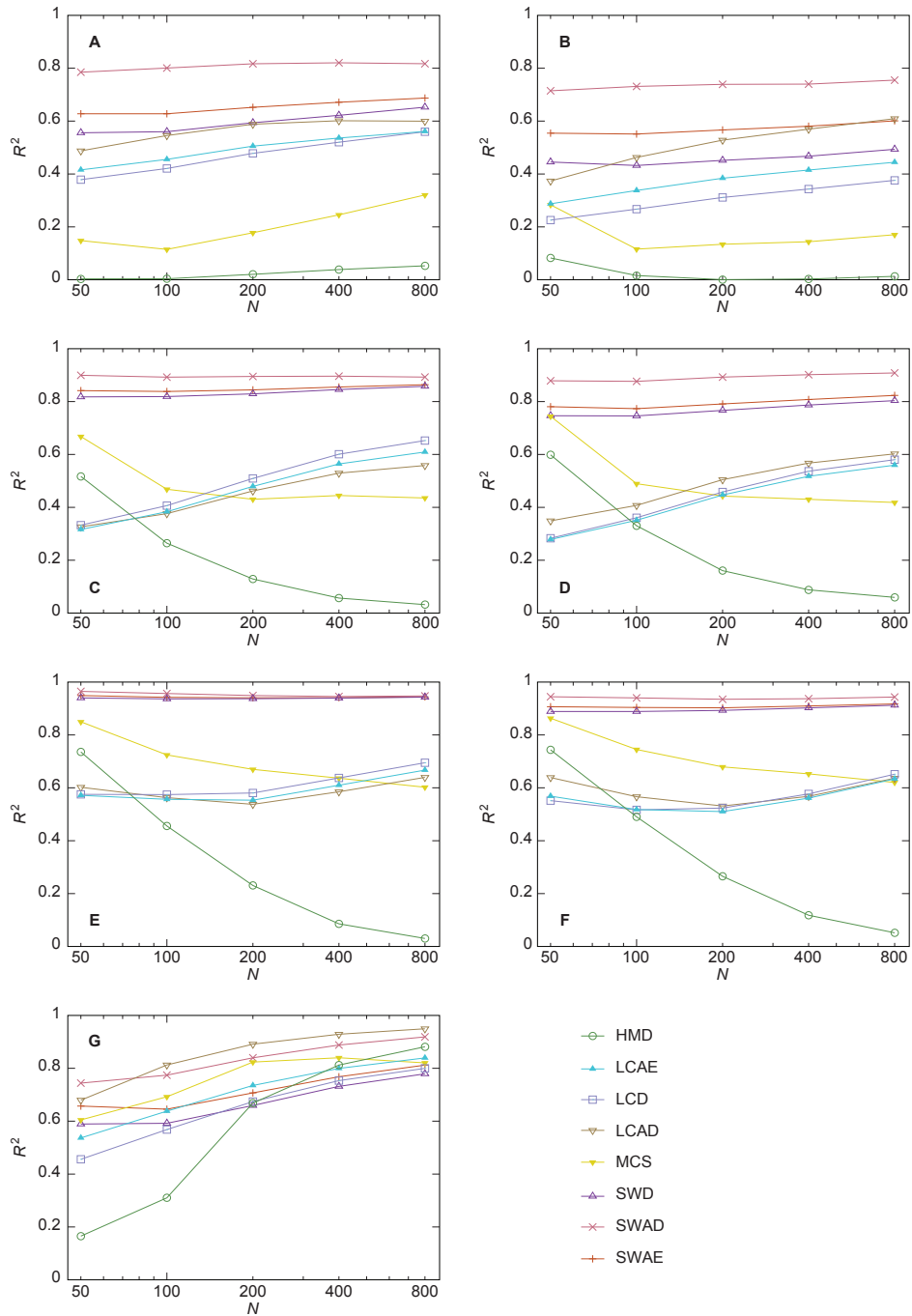

**Fig. 3.** $R^2$ values for the correlation between various distance metrics and maximal time to extinction as functions of number of vertices. Panels A and B show curves based on the configuration model with exponent −2; C and D show similar plots for exponent −2.5; E and F are corresponding plots for exponent −3; G shows the coefficients of determination for the Erdős-Rényi model. The abbreviations are explained in the Methods section. All data points are averaged over 1000 networks. Error bars would have been smaller than the symbol size.

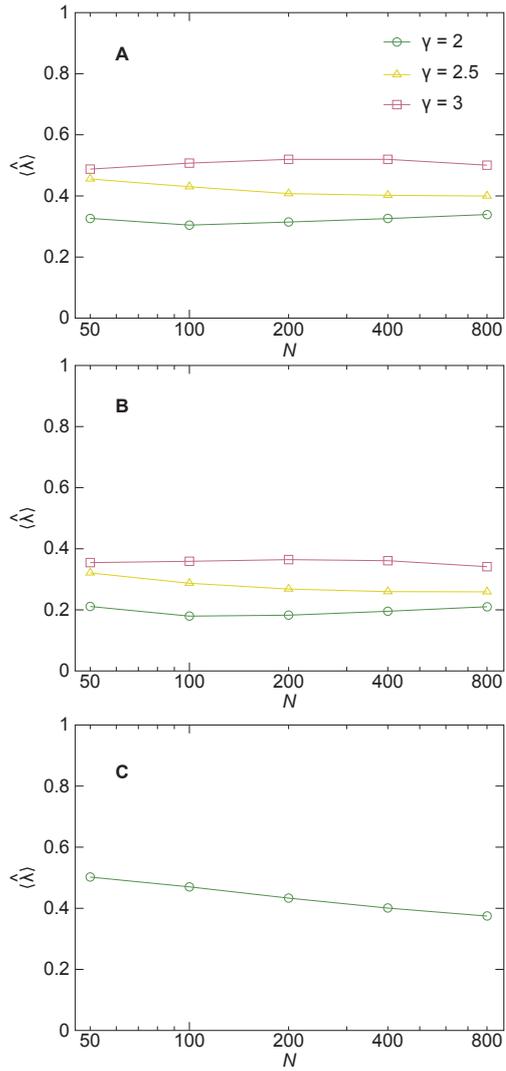

**Fig. 4.** The averages of λ-values maximizing the time to extinction, as functions of the system size. Panel A shows data for the configuration model; B is for the configuration model with added triangles; C is for the Erdős-Rényi model. All points are averaged over 1000 network realizations.